\documentstyle[referee,psfig,epsf]{l-aa}
\input psfig.tex
\def\lsim{\lower.5ex\hbox{$\; \buildrel < \over \sim \;$}}
\def\gsim{\lower.5ex\hbox{$\; \buildrel > \over \sim \;$}}
\newcommand{\eqb}{\begin{eqnarray}}
\newcommand{\eqe}{\end{eqnarray}}

\newbox\grsign \setbox\grsign=\hbox{$>$} \newdimen\grdimen \grdimen=\ht\grsign
\newbox\simlessbox \newbox\simgreatbox
\setbox\simgreatbox=\hbox{\raise.5ex\hbox{$>$}\llap
   {\lower.5ex\hbox{$\sim$}}}\ht1=\grdimen\dp1=0pt
\setbox\simlessbox=\hbox{\raise.5ex\hbox{$<$}\llap
   {\lower.5ex\hbox{$\sim$}}}\ht2=\grdimen\dp2=0pt
\def\gsim{\mathrel{\copy\simgreatbox}}
\def\lsim{\mathrel{\copy\simlessbox}}
 
\def\ref{\par\noindent\hangindent=1.5cm}
\begin{document}
\thesaurus{06(02.01.2, 02.02.1, 08.14.1, 02.08.1)}
%\title{Transonic Spherical Accretion in Pseudo-Scwarzschild Space-Time}
\title{Pseudo-Schwarzschild description of transonic spherical accretion 
onto compact objects}
\author{Tapas K. Das
\inst{1}
\and
                Aveek Sarkar \inst{2}
%S. Collin\inst{1}
%\and
%A.-M. Dumont\inst{1}
%\fnmsep\thanks{Just to show the usage
%       of the elements in the author field}
}

\offprints{Tapas K. Das}

\institute{Inter University
 Centre For Astronomy And Astrophysics, Post Bag 4 Ganeshkhind, Pune 411 007,
India\\
email: tapas@iucaa.ernet.in
\and
National Centre For Radio Astrophysics, TIFR, Post Bag 3 Ganeshkhind, Pune 411 007, India\\
email:
sarkar@ncra.tifr.res.in
%       \thanks{The university of heaven temporarily does not
%       accept e-mails}
}
\date{Received 6 March 2001/ Accepted 9 May 2001}
%; accepted ......\\
%\vskip0.05cm
%Astron. Astrophys. .......}
\maketitle
\markboth{Pseudo-Schwarzschild Spherical Accretion}{}

\maketitle

\begin{abstract}

A number of `modified' Newtonian potentials of various forms
are available in the literature
which accurately approximate some general relativistic effects important
for studying accretion discs around a Schwarzschild black hole. 
Such potentials may be called `pseudo-Schwarzschild' potentials because they nicely
mimic the space-time around a non-rotating/slowly 
rotating compact object.
In this paper, we examine the validity of the 
application of some of these potentials to study the spherically symmetric, 
transonic, hydrodynamic accretion onto a Schwarzschild black hole. 
By comparing the values of various dynamical and thermodynamic accretion parameters
obtained for flows using these potentials with full general relativistic 
calculations, we have shown that though the potentials
discussed in this paper were originally proposed to mimic the
relativistic effects manifested in disc accretion,
it is quite reasonable to use most of the
potentials in studying various dynamical as well as thermodynamic
quantities
for spherical accretion to compromise between the ease of
handling of a Newtonian description of gravity and the realistic situations
described by complicated general relativistic calculations.  Also we have shown that
depending on the chosen regions of parameter space
spanned by specific energy ${\cal E}$ and adiabatic index
$\gamma$ of the flow, one potential may have more
importance than another and we could identify which
potential is the best approximation for full general relativistic
flow in Scwarzschild space-time for particular
values of ${\cal E}$ and $\gamma$.

 \end{abstract}
\noindent
{\bf Published in the Astronomy and Astrophysics, 2001, v.374, p.1150-1160}.\\
\hrule

\section{Introduction}
Stationary, spherically symmetric and transonic hydrodynamic accretion of 
adiabatic fluid on to a gravitating astrophysical object at rest was 
studied in a seminal paper by Bondi (1952) using purely Newtonian Potential 
and by including the pressure effect of the accreting material. Later, 
Michel (1972) discussed fully general relativistic polytropic accretion on 
to a Schwarzschild black hole by formulating the governing equations for steady 
spherical flow of a perfect fluid in the Schwarzschild metric. Following 
Michel's relativistic generalization of Bondi's treatment,
 Begelman (1978) discussed some aspects of the critical (sonic) points of the flow
for such 
an accretion. Using an unrelaxed mono-energetic particle distribution and 
assuming the fact that the relaxation time of such a particle distribution 
is very long compared to the typical flow time scale or dynamical time 
scale of steady accretion on to black holes, Blumenthal and Mathews (1976) 
developed a model where the connection between the nonrelativistic to the 
relativistic regime of the spherically accreting material could be 
established. Taking the fully ionized one-temperature ($T_{electron}=T_{proton}$)
 hydrogen 
gas (governed by an exact relativistic equation of state) to be the 
fundamental constituent of the accreting material, Brinkmann (1980) treated 
spherically symmetric stationary accretion in Schwarzschild space time and 
showed that the temperature of accreting material at the Schwarzschild radius 
is one order of magnitude smaller than the flow temperature
obtained by using 
a simple polytropic equation of state.
Recently, Malec (1999) provided the solution for general
relativistic  spherical accretion with and without back reaction and showed
that relativistic effects enhance mass accretion when back reaction  is
neglected.\\
\noindent
Meanwhile, the theory of the 
accretion disc found prior importance because of the 
fact that disc accretion describes more realistic astrophysical situations 
found in nature. The beginning of modern accretion disc physics is 
traditionally attributed to the two classical articles by Shakura \& Sunyaev
(1973) and Novikov \& Thorne(1973) . While Shakura \& Sunyaev (1973) calculated 
the disk structure and related phenomena using purely Newtonian potential, Novikov 
\& Thorne provided a fully general relativistic description of accretion 
discs around black holes; later on, some aspects of which were modified by 
Riffert and Herold(1995). However, rigorous investigation of transonic disc 
structure was found to be extremely complicated in full general relativistic 
space time (Chakrabarti (1996) and references therein). At the same time it was 
understood that as relativistic effects play an important role in the 
regions close to the accreting black hole (where most of the 
gravitational potential energy is released), purely Newtonian gravitational 
potential (in the form ${\Phi}_{Newton}=-\frac{GM}{r}$) 
cannot be a realistic choice to describe 
transonic black hole accretion in general. To compromise between the ease of  
handling of a 
Newtonian description of gravity and the realistic situations 
described by complicated general relativistic calculations, a series of 
`modified' Newtonian potentials have been introduced 
to describe the general relativistic effects that are 
most important for accretion disk structure around Schwarzschild and Kerr 
black holes ( see Artemova et. al.(1996) for further discussion).
 Introduction of such potentials allows one to investigate the 
complicated physical processes taking place in disc accretion in a
semi-Newtonian framework by avoiding pure general relativistic calculations
that 
most of the features of spacetime around a compact object are retained and
some crucial properties of the analogous relativistic 
solutions of disc structure could be reproduced with high accuracy.
Hence, those potentials might be designated as `pseudo-Kerr' or `pseudo- 
Schwarzschild' potentials, depending on whether they are used to mimic the 
space time around a rapidly rotating or non rotating/ slowly rotating
(Kerr parameter $a\sim0$) black 
hole respectively.\\
\noindent
It is important to note that although a number of such 
`pseudo' potentials are available in the 
literature to study various aspects of 
disc accretion, no such potentials are available which had been solely 
derived to describe spherically symmetric accretion on to a Schwarzschild 
(or Kerr) black hole. In this paper, we will concentrate on some of the 
`pseudo-Schwarzschild' {\it disc} potentials (potentials introduced to study 
{\it accretion discs}
around a Schwarzschild black hole) to investigate whether those potentials could 
be used to study Spherical accretion, and if 
so, how `good' the choice would be for various such potentials.
Also, we would like to check which potential among those would be the `best-fit' to 
approximate the full general relativistic description of
transonic, spherically symmetric accretion on to a Schwarzschild black 
hole. In doing so, we 
solve the equations of motion of spherically accreting fluid in full
Schwarzschild space-time as well as for
motion under various `pseudo'-potentials,
to study the variation of different 
dynamical and thermodynamic quantities (like Mach number of the flow, flow 
temperature  etc.) with radial distance measured from the accreting black hole for 
the {\it full
general relativistic spherical flow} (hereafter FGRSF)
 as well as for accretion using 
various `pseudo- Schwarzschild' potentials. We then compare the results 
obtained using such potentials with the solutions of exact relativistic 
problems in a Schwarzschild metric. The plan of the paper is as follows:
In next section, we will describe four `pseudo-Schwarzschild' disc potentials 
available in the literature and some of their basic features. In \S 3, 
we will provide the basic equations governing spherically symmetric 
accretion in full relativistic as well as in various 
`pseudo'--relativistic spacetimes.  In \S 4 we 
will discuss  how to solve those equations to find various
dynamical quantities which are to be mutually compared and
we present our results. Finally in \S 5 we conclude by discussing
the suitability of various `pseudo' potentials 
in approximating the results obtained from
exact relativistic 
calculations. For the rest of this paper, we will use the terms `modified-
Newtonian potential' and `pseudo (Schwarzschild) potentials'
synonymously.
\section {Some basic features of various `pseudo-Schwarzschild' potentials}
From now, we will define the Schwarzschild radius $r_g$ as 
$$
r_g=\frac{2G{M_{BH}}}{c^2}
$$
(where  $M_{BH}$  is the mass of the black hole, $G$
is universal gravitational 
constant and $c$ is velocity of light in vacuum) so that the marginally bound 
circular orbit $r_b$ and the last stable circular orbit $r_s$
take the values $2r_g$
and $3r_g$ respectively for a typical Schwarzschild black hole. Also,
total 
mechanical energy per unit mass on $r_s$ (sometimes called 
`efficiency' $e$) may be computed as $-0.057$ for this 
case. Also, we will use a simplified geometric unit throughout this paper where 
radial distance $r$ is be scaled in units of $r_g$, radial dynamical 
velocity 
$u$ and polytropic sound speed $a$ of
the flow is scaled in units of $c$ (the 
velocity 
of light in vacuum), mass $m$ is scaled in units of $M_{BH}$
and all 
\begin{figure}
\vbox{
\vskip -5.8cm
\centerline{
\psfig{file=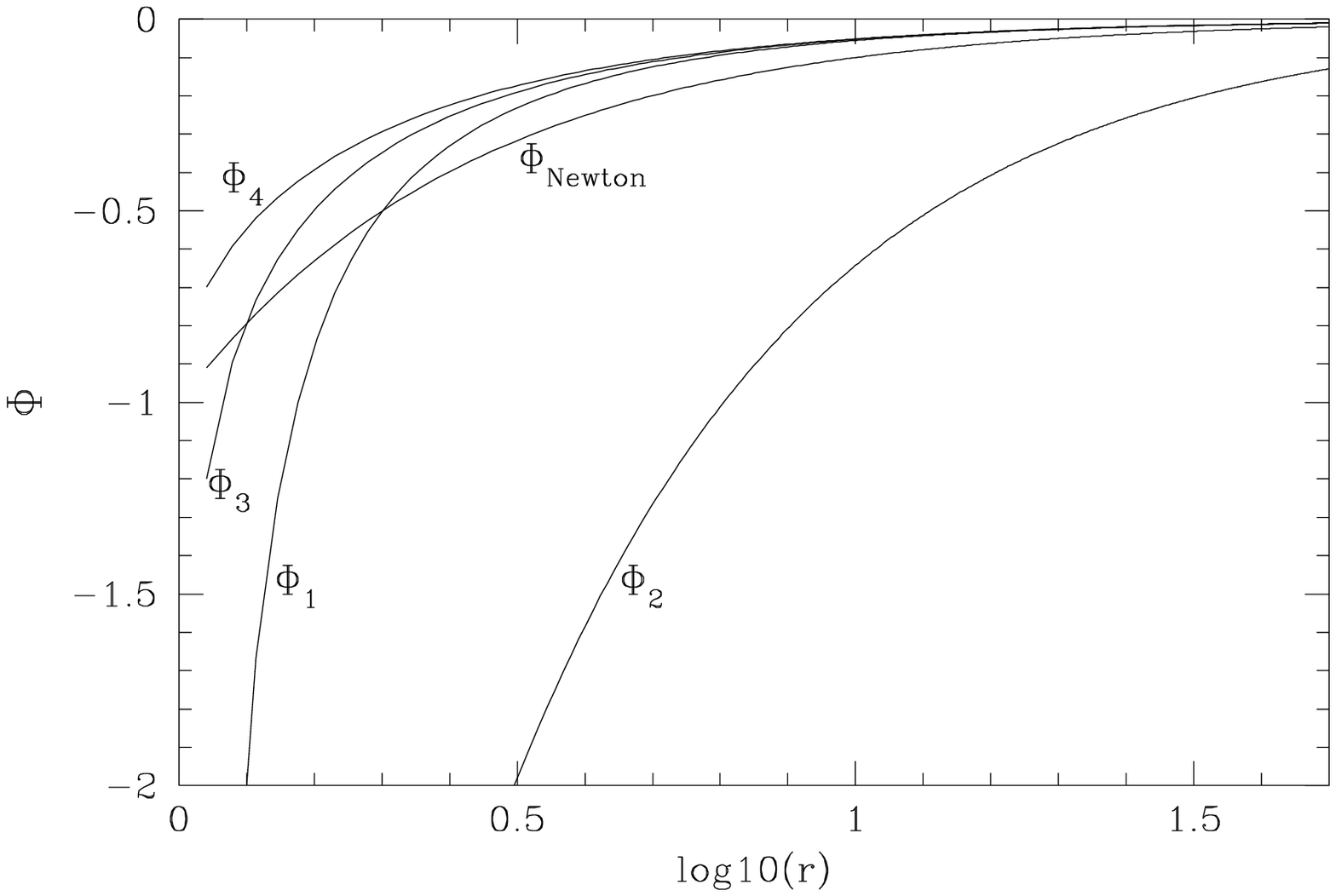,height=15cm,width=15cm}}}
\noindent {{\bf Fig. 1:}
Newtonian potential $\Phi_{Newton}(r)$ and other
pseudo-potentials $\Phi_i(r)$'s ($i=1,2,3,4$) are plotted as a function of
the logarithmic radial distance from the accreting black hole. Except $\Phi_2$, all
other pseudo-potentials have a singularity at one Schwarzschild radius
$r_g$. For $r~>~2r_g$, while the relative stiffness factor ${\cal S}$ is maximum
for $\Phi_4$, it is minimum for $\Phi_2$, see text for details.}
\end{figure}
other derived quantities would be scaled 
accordingly. Also, for simplicity,we will use $G=c=M=1$.
Below we would like to 
briefly describe four 
different `pseudo- Schwarzschild' potentials (expressed in 
the system of units discussed above) and to provide the `free fall' 
acceleration obtained using such potentials in compact form.\\
\noindent
Paczy\'nski and Wiita (1980) introduced a `pseudo-schwarzschild'
potential of the form
$$
\Phi_{1}=-\frac{1}{2(r-1)}
\eqno{(1a)}
$$
which accurately reproduces the positions of $r_s$ and $r_b$ 
and gives the value 
of efficiency to be $-0.0625$. Also the Keplarian distribution of angular 
momentum obtained using this potential is exactly same as 
that obtained in pure 
Schwarzschild geometry. Although this potential (as well as the other
`pseudo' potentials available in the literature) does not 
satisfy the boundary condition exactly on the horizon, however,
 close to the 
horizon, accretion flows are supposed to be highly supersonic and the 
dynamical infall time scale becomes too small to allow significant radiation 
from infalling fluid. Even if there are significant amounts
of radiation, it 
does not contribute too much to the overall radiation coming out from the 
disc, due to the fact that radiation emitted from the 
very close viscinity of the 
black hole is highly redshifted. Thus, errors in calculation
close to the horizon may 
have very little impact on the emitted disc spectrum and for all practical 
purposes, the 
Paczy\'nski and Wiita (1980)
potential is considered to be the best approximation of 
pure Schwarzschild spacetime while treating the disc accretion 
(Artemova 1996) . It is worth mentioning here that this potential 
was first introduced to study a thick accretion disc with super Eddington 
Luminosity. Also,
it is interesting to note that although it had been thought of
in 
terms of disc accretion, it is spherically symmetric (with a scale shift of 
$r_g$). So in principle, it can definitely be used to study spherical 
accretion also on to a Schwarzschild black hole.\\
\noindent
To analyse the normal modes of accoustic oscillations within a 
thin accretion 
disc around a compact object (slowly rotating black hole or weakly 
magnetized neutron star), Nowak and Wagoner (1991) approximated some of the 
dominant relativistic effects of the accreting 
black hole (slowly rotating or 
nonrotating) via a modified Newtonian potential of the form
$$
\Phi_{2}=-\frac{1}{2r}\left[1-\frac{3}{2r}+12{\left(\frac{1}{2r}\right)}^2\right
]
\eqno{(1b)}
$$
$\Phi_2$ has correct form of $r_s$ as in the Schwarzschild case
but is unable to 
reproduce the value of $r_b$.
 This potential has the correct general relativistic value of the
angular velocity (as measured at infinity) at $r_s$. Also it reproduces the
radial epicyclic frequency $\kappa$ (for $r>r_s$) close to its value obtained
from general relativistic calculations. However, this potential gives the
value of efficiency as $-0.064$ which is larger than that produced by 
$\Phi_1$, hence the disc spectrum computed using $\Phi_2$ would be more 
luminous compared to a disc structure studied using $\Phi_1$.\\
\noindent
Remembering  that the free-fall acceleration plays a very crucial 
role in Newtonian gravity, Artemova et. al. (1996) proposed two different 
`pseudo' potentials to study disc accretion around a non-rotating black hole.
The first potential proposed by them produces exactly the
same value of the free-fall
acceleration of a test particle at a given value of $r$ as is obtained
for a test particle at rest with respect to the Schwarzschild reference
frame, and is given by
$$
\Phi_{3}=-1+{\left(1-\frac{1}{r}\right)}^{\frac{1}{2}}
\eqno{(1c)}
$$
The second one gives the value of the free fall acceleration that is equal 
to the value of the covariant component of the three dimensional free-fall 
acceleration vector of a test particle that is at rest in the Schwarzschild 
reference frame and is given by
$$
\Phi_{4}=\frac{1}{2}ln{\left(1-\frac{1}{r}\right)}
\eqno{(1d)}
$$
Efficiencies produced by $\Phi_3$ and $\Phi_4$ are $-0.081$ and $-0.078$ 
respectively.The magnitude of efficiency produced by $\Phi_3$
being 
maximum,calculation of disc structure using $\Phi_3$
will give  the maximum 
amount of energy dissipation and the corresponding spectrum would be the 
most luminous one. However, as both $\Phi_3$ and 
$\Phi_4$ stems from the 
consideration of free fall acceleration and calculates the dependence of 
free fall accleration on radial distance in the Schwarzschild metric 
(which 
describes a spherically symmetric gravitational field in vacuum), it 
appears to be 
quite justified to use those potentials to study spherically symmetric 
accretion.\\
\noindent
From now we will refer to 
all these four potentials as $\Phi_i$ in 
general where $\left\{i=1,2,3,4\right\}$ would correspond to $\Phi_1$
(eqn. 1(a)), $\Phi_2$ (eqn. 1(b)), $\Phi_3$ (eqn. 1(c)) and $\Phi_4$ (eqn. 1(d))
respectively.
In Fig. 1, we plot various $\Phi_i$'s as a function of the radial distance
measured from the accreting black hole in units of $r_g$. Also in the same plot,
purely Newtonian potential $\Phi_{Newton}$ is plotted.
If we now define a quantity ${\cal S}_i$ to be the `relative stiffness'
of a potential $\Phi_i$ as:
$$
{\cal S}_i=\frac{\Phi_i}{r}
$$
(that is, ${\cal S}_i$ is a measure of the numerical value of any $i$th 
potential at a radial distance $r$), we find that for $r~>~2r_g$:
$$
{\cal S}_2~<~{\cal S}_{Newton}~<~{\cal S}_1~<~{\cal S}_3~<~{\cal S}_4
$$
which indicates that while $\Phi_2$ is a `flatter' potential compared to the
pure Newtonian potential $\Phi_{Newton}$, all other `pseudo' potentials are
`steeper' to  $\Phi_{Newton}$ for $r~>~2r_g$. \\
\noindent
One can write the modulus of free fall 
acceleration obtained from all `pseudo' potentials except for $\Phi_2$ 
in a compact form as
$$
\left|{{{{{\Phi}^{'}}_{i}}}}\right|=\frac{1}{2{r^{2-{\delta}_{i}}
{\left(r-1\right)}^{\delta_{i}}}}
\eqno{(2a)}
$$
where ${\delta_{1}}=2$, $\delta_3=\frac{1}{2}$ and $\delta_4=1$. 
$\left|{{{{{\Phi}^{'}}_{i}}}}\right|$
denotes the absolute value of the 
{\it space derivative} of $\Phi_i$, i.e.,
$$
\left|{{{{{\Phi}^{'}}_{i}}}}\right|=\left|{\frac{d{\Phi_i}}{dr}}\right|
$$
whereas acceleration produced by $\Phi_2$ can be computed as,
$$
{\Phi_2}^{'}=\frac{1}{2r^2}\left(1-\frac{3}{r}+\frac{9}{2r^2}\right)
\eqno{(2b)}
$$
In the next section,we would like to describe how one can investigate 
transonic spherical accretion using these potentials. Also, we will discuss 
how to calculate various dynamical quantities for full general relativistic 
bondi flow in a Schwarzschild metric. One standard method to investigate 
classical transonic bondi flow is to formulate the basic conservation 
equations, i.e.,conservation of baryon number (obtained by integrating
continuity 
equation) and conservation of specific energy (obtained by
integrating Euler's equation 
using a specific equation of state for accreting matter) and then to 
simultaneously  solve these conservation equations to get 
critical (sonic) quantities as functions of various accretion parameters 
(like specific energy ${\cal E}$, adiabatic index $\gamma$
or  accretion rate ${\dot M}_{in}$  in the flow)
and also to calculate the values of various dynamical and 
thermodynamic quantities (like Mach number $M$, of the 
flow, flow temperature, $T$, etc.) as functions of various accretion 
parameters (like ${\cal E}, \gamma, 
{\dot M}_{in}$ etc.) or radial distance (measured from the central 
accretor in the unit of Schwarzschild radius $r_g$). 
Also a common practice is 
to study the variation of 
Mach number $M$ (ratio of the local dynamical velocity 
$u$
to the local sound velocity $a$; $M=\frac{u}{a}$) 
with radial distance $r$ (measured in units of $r_g$) to 
investigate the `transonicity' of the 
flow for a fixed set of input parameters.
We will 
perform the above mentioned calculations for accretion in pure Schwarzschild 
spacetime as well as for accretion in `pseudo-schwarzschild' space time 
using $\Phi_i$'s and compare the results obtained for various 
$\Phi_i$'s with exact 
relativistic calculations.
As accretion onto a 
black hole is {\it necessarily} transonic to satisfy
the inner boundary condition at the event horizon, unlike Bondi's 
(1952) original work,
here we will concentrate only on transonic solutions, i.e., we will deal with accretion
(wind) which is subsonic (supersonic) far away from the black hole and 
approaches (moves away from)
the hole supersonically (subsonically)
after crossing a sonic point (the location of which can be 
determined as a function of ${\cal E}$ and $\gamma$, see \S 3) on its way 
towards (away from) the accretor.
\section{Governing equations}
\subsection{Full General Relativistic Spherical Flow (FGRSF) in
Schwarzschild space-time}
For a schwarzschild metric of the form 
$$
ds^2=dt^2\left(1-\frac{1}{r}\right)-dr^2{\left(1-\frac{1}{r}\right)}^{-1}-r^2{\left(d{\theta}^2+sin^2{\theta}{d\phi^2}\right)}
$$
the energy momentum tensor $T^{{\alpha}{\beta}}$ for a perfect fluid
can be written as (Shapiro \& Teukolsky 1983)
$$
T^{\alpha\beta}={\epsilon}u^{\alpha}u^{\beta}+p\left(u^{\alpha}u^{\beta}-
g^{\alpha\beta}\right)
$$
where ${\epsilon}$ and $p$ are proper energy density and pressure of the 
fluid (evaluated in the local inertial rest frame of the fluid)
respectively and $u^{\alpha}$ is the four velocity commonly known as
$$
u^{\alpha}=\frac{dx^{\alpha}}{ds}
$$
Equations of motion which are to be solved for our purpose are,\\
1) Conservation of mass flux or baryon number conservation:
$$
{\left({\rho}{u_{\alpha}}\right)}_{;~{\alpha}}=0
\eqno{(3a)}
$$
and \\
2) Conservation of momentum or energy flux (the 
general relativistic Euler 
equation obtained by taking the four divergence of $T^{{\alpha}{\beta}}$):
$$
\left({\epsilon}+p\right){u_{{\alpha};{\beta}}}u^{\beta}=
-p_{,\alpha}-u_{\alpha}p_{,\beta}u^\beta_{,}
\eqno{(3b)}
$$
where the semicolons denote the covariant derivatives. \\
\noindent
Following Michel (1972), one can rewrite eqn. 3(a) and eqn. 3(b) for 
spherical accretion as 
$$
4{\pi}{\rho}ur^2={\dot M}_{in}
\eqno{(4a)}
$$
and
$$
\left(\frac{p+\epsilon}{\rho}\right)^2\left(1-\frac{1}{r}+u^2\right)={\bf C}
\eqno{(4b)}
$$
as two fundamental conservation equations for time
independant hydrodynamical flow of matter on to a
Schwarzschild black hole without back-reaction of the
flow on to the metric itself. ${\dot M}_{in}$   is the mass
accretion rate and ${\bf C}$ is a constant 
(related to the total enthalpy influx)
to be evaluated
for a specific equation of state.\\
\noindent
For a polytropic equation of state i.e,
$$
p=K{\rho}^{\gamma}
$$
and defining ${\dot {\cal M}}={\dot M}_{in}{\gamma}^nK^n$,
where ${\dot {\cal M}}$ is a measure of the 
entropy of the flow ($n$ is a polytropic constant of the flow defined as
$n=\left(\gamma-1\right)^{-1}$)
and is another
conserved quantity of the flow, one can rewrite
conservation equations 4(a) and 4(b) as (Chakrabarti (1996) and
references therein):
$$
{\cal E}=1+hu_t=1+\left(\frac{p+\epsilon}{\rho}\right)
{\left(\frac{1-\frac{1}{r}}{1-u^2}\right)}^{\frac{1}{2}}
\eqno{(5a)}
$$
and
$$
{\dot {\cal M}}=4{\pi}{\left(\frac{a^2}{1-na^2}\right)^n}uu_tr^2
\eqno{(5b)}
$$
Where ${\cal E}, h,u_t$  and $a$ are the conserved specific energy
of the flow {\it excluding} its rest mass,
specific enthalpy, specific binding
energy and local adiabatic sound speed
respectively. Eq. 5(b) may be considered as the outcome of
the conservation
of mass and entropy along the flow line.
The expression for $a$ can be written as (Weinberg (1972), Frank et. al.
(1992)):
$$
a={\left(\frac{{\partial}p}{{\partial}\rho}\right)}^{\frac{1}{2}}
_{Constant~Specific~Entropy}=\sqrt{\frac{{\gamma}{p}}{\rho}}=
\sqrt{\frac{{\gamma}{\kappa}{T}}{{\mu}{m_H}}}
\eqno{(6)}
$$
Where $T$ is the flow temperature, $\mu$ is the mean molecular weight and
$m_H{\sim}m_p$ is the mass of the hydrogen atom.\\
\noindent
One can now easily derive the expression for the velocity
gradient $\left(\frac{du}{dr}\right)$ (by differentiating eqn. 5(a) and 5(b))
as 
$$
\frac{du}{dr}=\frac{u\left(1-u^2\right)\left\{a^2\left(4r-3\right)-1\right\}}
{2r\left(r-1\right)\left(u^2-a^2\right)}
\eqno{(7a)}
$$
Since the flow is assumed to be smooth everywhere, if
the denominator of eqn. 7(a)  vanishes at any radial distance
$r$, the numerator must also vanish there to maintain the
continuity of the flow. One therefore arrives at the so
called `sonic point (alternately, the `critical point'
\footnote{Hereafter, we will use `critical ponts' and `sonic points'
synonymously.}) conditions'  by simultaneously making
the numerator and denominator of eqn. 7(a) equal zero.
The sonic point conditions then can be expressed as follows
$$
u_c=a_c=\frac{1}{4r_c-3}
\eqno{(7b)}
$$
For a specific value of ${\cal E}$ and $\gamma$,
location of the sonic point $r^{gr}_c$
can be obtained by solving the following equation
algebraically
$$
{\cal E}-\frac{\left(\gamma-1\right)\left(4r_c-3\right)}
{4r_c\left(\gamma-1\right)-
\left(3{\gamma}+4\right)}\sqrt{1-\frac{3}{4r_c}}+1=0
\eqno{(7c)}
$$
The spherical surface of radius $r=r^{gr}_c$ can be defined as the
`accoustic horizon' because for $r~<~r^{gr}_c,~u~>~a$  and any
accoustic disturbances created in this region are
advected towards the black hole. Thus no accoustic
disturbances created within this radius can cross the
accoustic horizon and escape to the region $r~>~r_c$. \\
\noindent
To
determine the behaviour of the solution near the sonic
point, one needs to evaluate the value of $\left(\frac{du}{dr}\right)$
at that
point (we denote it by $\left(\frac{du}{dr}\right)_c$)
by applying L 'Hospitals' rule to eqn. 7(a). It is
easy to show that $\left(\frac{du}{dr}\right)_c$
can be obtained by solving the
following quadratic equations algebraically:
$$
\left(\frac{du}{dr}\right)^2_c+\frac{\left(\gamma-1\right)
\left(16r^2_c-16r_c-8{\gamma}r_c+6\gamma+3\right)}
{3r_c\left(4r_c-3\right)^{\frac{3}{2}}}
\left(\frac{du}{dr}\right)_c
$$
$$
+
\frac{\left(\gamma-1\right)\left(2r_c-1\right)\left(24r^2_c-28r_c-8r^2_c\gamma+4r_c\gamma+3\gamma+6\right)}
{2r_c\left(4r_c-3\right)^{\frac{3}{2}}\left(4r_c-3\right)\left(r_c-1\right)}
=0
\eqno{(7d)}
$$
It is now quite straightforward to simultaneously
solve eqn. 5(a) and eqn. 5(b) to
get the integral curves of the flow for a fixed value
of ${\cal E}$ and $\gamma$. Detailed methodology for this purpose will 
be discussed
in \S 4.
\subsection{Spherical flow in various
Pseudo-Schwarzschild space time}
For any `pseudo-Schwarzschild' potential $\Phi_i$, the
equation of motion for spherically accreting  material
onto the accretor
is given by 
$$
\frac{{\partial{u}}}{{\partial{t}}}+u\frac{{\partial{u}}}{{\partial{r}}}+\frac{1}{\rho}
\frac{{\partial}P}{{\partial}r}+{\Phi_i}^{'}=0
\eqno{(8a)}
$$
where symbols have their
usual meaning.
The first term of eqn. 8(a) is the Eulerian time derivative of the
dynamical velocity at a given $r$, the second term 
is the `advective' term, the third term 
is the
momentum deposition due to pressure gradient and the
final term is due to the gravitational acceleration
for a particular $i$th potential $\Phi_i$. The continuity
equation can be written as 
$$
\frac{{\partial}{\rho}}{{\partial}t}+\frac{1}{r^2}\frac{{\partial}}{{\partial}r}\left({\rho}ur^2\right)=0
\eqno{(8b)}
$$
For a polytropic equation of state, the steady state
solution (apart from a geometric factor of $4\pi$) 
of eqn. 8(a) and eqn. 8(b) is\\
1) Conservation of specific energy ${\cal E}$ of the flow:
$$
{\cal E}=\frac{u^2}{2}+\frac{a^2}{{\gamma}-1}+\Phi_i
\eqno{(9a)}
$$
and\\ \\
2) Conservation of Baryon number (or accretion rate ${\dot M}_{in}$):
$$
{\dot M}_{in}=4{\pi}{\rho}ur^2
\eqno{(9b)}
$$
Using ${\dot {\cal M}}$  as defined earlier
(${\dot {\cal M}}={\dot M}_{in}{\gamma}^nK^n$), 
eqn. 9(b) can be rewritten as 
$$
{\dot {\cal M}}=4{\pi}a^{2n}ur^2
\eqno{(9c)}
$$
It is now quite straightforward to derive the space
gradient of dynamical velocity $\left(\frac{du}{dr}\right)_i$ 
for flow in any particular $i$th potential $\Phi_i$ as
$$
{\left(\frac{du}{dr}\right)}_i=\frac{\frac{2a^2}{r}-
{\Phi_i}^{'}}{u-\frac{a^2}{u}}
\eqno{(10a)}
$$
Using either eqn. 2(a) or eqn. 2(b), one can substitute the
value of any specific $i$th ${\Phi_i}^{'}$ in eqn. 10(a) to get the
value of ${\left(\frac{du}{dr}\right)}_i$
for motion under particular $i$th
pseudo-Schwarzschild potential. One can also calculate
the sonic point quantities (as described in \S 3.1)
as
$$
{u^i}_c={a^i}_c=\sqrt{\frac{r^i_c}{2}{{\Phi_i}^{'}}{{\Bigg{\vert}}}_c}
\eqno{(10b)}
$$
where superscript $i$ denotes the specific
value of sonic quantities for a particular $i$th
potential $\Phi_i$, and ${\Phi^{'}_i}{\bigg{\vert}}_c$ is the 
value of
$\left(\frac{d{\Phi_i}}{dr}\right)$ evaluated at the corresponding sonic point
$r^i_c$. The value of sonic point $r^i_c$ for any $i$th potential 
$\Phi_i$ can be obtained
by algebraically solving the following equation
$$
{\cal E}-\frac{1}{2}\left(\frac{\gamma+1}{\gamma-1}\right)r^i_c
{{\Phi_i}^{'}}{{\Bigg{\vert}}_c}
-{\Phi_i}{\Bigg{\vert}}_c=0
\eqno{(10c)}
$$
where ${\Phi_i}{\bigg{\vert}}_c$ is the value of $i$th potential at the
corresponding sonic point $r^i_c$. 
 Similarly, the value of $\left(\frac{du}{dr}\right)_i$ for any $\Phi_i$
at its
corresponding sonic point $r^i_c$ can be obtained by
solving the following quadratic equation:
$$
\left(1+\gamma\right){{\left(\frac{du}{dr}\right)}^2}_{c_,i}+
2.829\left(\gamma-1\right)\sqrt\frac{{{\Phi_i}^{'}}
{{\Bigg{\vert}}_c}}{r^{i}_c}{\left(\frac{du}{dr}\right)}_{c_,i}
+\left(2{\gamma}-1\right)
\frac{
{{\Phi_i}^{'}{\Bigg{\vert}}_c}}
%{{r^i_c}}
{{r^i_c}}
+{{\Phi_i}^{''}}{{\Bigg{\vert}}_c}=0
\eqno{(10d)}
$$
where ${{\Phi_i}^{''}}{{\Bigg{\vert}}_c}$
is the value of $\frac{d^2\Phi_i}{dr^2}$  
at the corresponding
critical point $r^i_c$.\\
\noindent
One can simultaneously solve eqn. 9(a) and eqn. 9(b)
(alternatively, eqn. 9(a) and eqn. 9(c))
for any specific $\Phi_i$ for a
fixed value of ${\cal E}$ and $\gamma$
to obtain various dynamical and thermodynamic flow quantities. 
We discuss the detailed
methodology in \S 4.
\begin{figure}
\vbox{
\vskip -0.0cm
\centerline{
\psfig{file=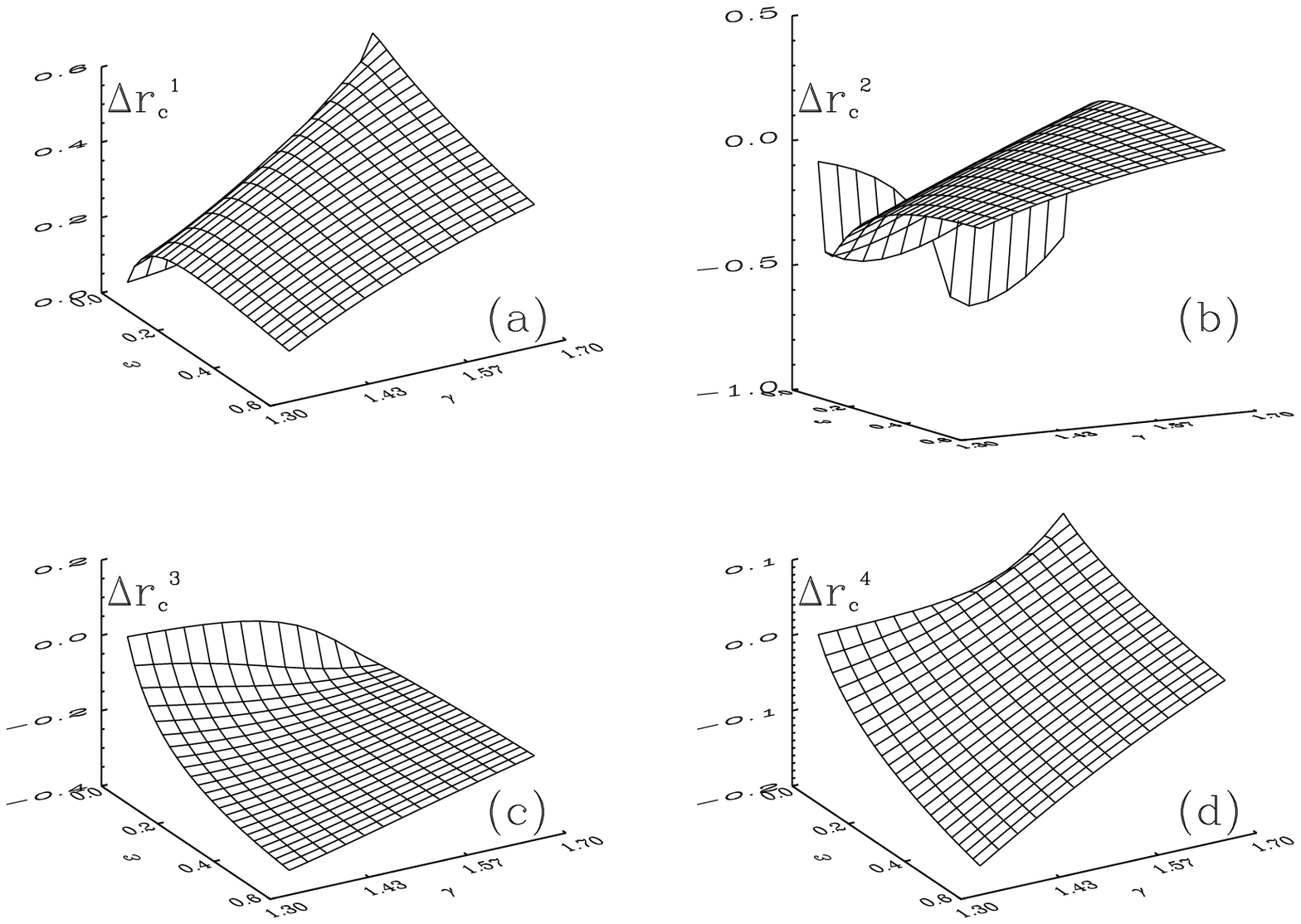,height=17cm,width=19cm}}}
\noindent {{\bf Fig. 2 (a,b,c,d):}
The deviation
(${\Delta}r^i_c$) of the value of sonic points
$r^i_c$'s obtained for flows using various $\Phi_i$'s from $r^{gr}_c$
as a function of specific energy ${\cal E}$ and adiabatic
index $\gamma$
of the flow is plotted for flows in $\Phi_1$ (Fig. 1(a)), $\Phi_2$
(Fig. 1(b)), $\Phi_3$ (Fig. 1(c)) and $\Phi_4$ (Fig. 1(d)). It is observed
that while ${\Delta}r^1_c$ is a minimum for ultra-relativistic ($\gamma=
\frac{4}{3}$) flow, ${\Delta}r^4_c$ becomes a minimum as the flow approaches
a pure-non-relativistic ($\gamma=\frac{5}{3}$) limit, see text for
detail.}
\end{figure}
\section{Solution procedure and results}
\subsection{Sonic points as a function of ${\cal E}$ and $\gamma$}
\noindent
Solving eqn. 7(c) and eqn. 10(c) for a fixed value of ${\cal E}$ and
$\gamma$, the sonic point could be obtained for FGRSF and for flows in various 
pseudo-potentials respectively. We denote $r^{gr}_c$ to 
be the sonic point for
FGRSF and $r^i_c$ to be the sonic point for flow in $i$th potential 
$\Phi_i$, and define ${\Delta}r^i_c$ as:
$$
{\Delta}r^i_c=\frac{r^i_c-r^{gr}_c}{r^{gr}_c}
\eqno{(11)}
$$
In Fig. 2., we plot various ${\Delta}r^i_c$'s 
(along the $z$ axis) as a function of specific energy
${\cal E}$ (along the $x$ axis) and adiabatic index $\gamma$ (along the
$y$ axis) of the flow. The difference in the general profile
of variation of ${\Delta}r^i_c$ for various $\Phi_i$'s
with ${\cal E}$ and $\gamma$ is quite obvious from the figure. 
 It is observed that when $\gamma$ is kept fixed at its minimum value 
($\gamma=\frac{4}{3}$), ${\Delta}r^1_c$ (${\Delta}r_c$ for flows in
$\Phi_1$) is always positive (which means $r^1_c~>~r^{gr}_c$ always) and
is at a 
minimum for low energy and starts increasing nonlinearly with an
increase of
${\cal E}$. At a certain value of ${\cal E}$, it produces a peak and starts falling,
with further increase of ${\cal E}$. The peak value of ${\Delta}r^1_c$ is observed
to be $\sim0.15$. Unlike ${\Delta}r^1_c$, 
${\Delta}r^2_c$ falls very sharply with ${\cal E}$
and shows a `dip' at ${\cal E}\sim0.04$, then starts increasing with ${\cal E}$
monotonically and nonlinearly,
keeping ${\Delta}r^2_c$ negative for all values
of ${\cal E}$ which indicates that $r^2_c~<~r^{gr}_c$ 
for ultra-relativistic flow ($\gamma=\frac{4}{3}$),
\footnote{{Hereafter, we will describe the flow to be ultra-relativistic
for $\gamma=\frac{4}{3}$ and purely non-relativistic for $\gamma=\frac{5}{3}$
according to standard practice (Frank. et. al (1992)).}}
for all values of ${\cal E}$ we consider here.
For $\Phi_3$ and $\Phi_4$, we observe that 
${\Delta}r^3_c$
and ${\Delta}r^4_c$ monotonically decreases
with ${\cal E}$ and also ${\Delta}r^3_c$ 
and ${\Delta}r^4_c$ are always negative e.g., $r^3_c,r^4_c ~<~r^{gr}_c$ for all 
${\cal E}$ we consider here. It is observed that 
for ultra-relativistic flow, 
$\Phi_1$ is the best approximation to study transonic
spherically symmetric polytropic accretion to produce the sonic point
closest to the sonic point formed for FGRSF in general, especially
in {\it high energy regime},
i.e., for high value of ${\cal E}$, while $\Phi_2$ is the worst approximation
for studying the same phenomena for same set of boundary conditions. If we 
define a hypothetical quantity ${\cal G}^i_{rc}$ which is a measure of `goodness' 
of a particular $i$th potential $\Phi_i$ regarding the closest approximation
of 
sonic points obtained for
FGRFS, we find that following sequence holds good for
ultra-relativistic flow :
$$
{ \bf {\cal {G}}}^1_{rc}~{{>}}~{ \bf {\cal {G}}}^4_{rc}~{{>}}~{ \bf {\cal {G}}}^
3_{rc}~{{>}}
~{ \bf {\cal {G}}}^2_{rc}
\eqno{(12)}
$$
The situation starts changing as the value of $\gamma$ increases. 
The general profile of ${\Delta}r^1_c$ is unchanged but its peak 
starts shifting towards the lower ${\cal E}$ values with increasing 
$\gamma$ and the overall deviation starts increasing. For 
purely nonrelativistic flow ($\gamma=\frac{5}{3}$),
${\Delta}r^1_c{\bigg{\vert}}_{max}$ is found to be 
quite large. For all 
values of $\gamma$, ${\Delta}r^1_c$ is found to be positive,
which means 
that $r^1_c~>~r^{gr}_c$ for {\it all} values of $\gamma$ and ${\cal E}$.\\
\noindent
However, the behaviour of ${\Delta}r^2_c$ starts changing drastically with 
an increase in $\gamma$.
 The `dip' produced by ${\Delta}r^2_c$ for low values of $\gamma$ starts
smearing out with an increase in 
$\gamma$ and ${\Delta}r^2_c$
becomes positive with higher ${\cal E}$ for a fixed value of $\gamma$.
 For higher $\gamma$ it has been observed that if we study variation of
${\Delta}r^2_c$  with ${\cal E}$ and $\gamma$, ${\Delta}r^2_c$ 
starts with a very high negative value at lower ${\cal E}$ 
with $\left|{\Delta}r^2_c\right|i_{max}$ 
(where $\left|{\Delta}r^2_c\right|_{max}$ stands for the maximum value
of the {\it modulus} of ${\Delta}r^i_c$),
monotonically decreasing with
$\gamma$, 
reaches zero and becomes positive and increases nonlinearly up to 
 a maximum value (this time, a `peak') and then starts falling
again with an increase of ${\cal E}$. So for high values of
$\gamma$, smearing out of the `dip' in ${\Delta}r^2_c$  vs ${\cal E}$ 
curve is
compensated by the appearance of 
a `peak' at some value of ${\cal E}$. It is also observed that 
the peak starts shifting towards the higher value of ${\cal E}$ as
$\gamma$ is increased. It is also observed that as we go from ultra-relativistic
flow towards purely non-relativistic flow, 
a zero appears in ${\Delta}r^2_c$  vs ${\cal E}$
curve, which indicates that at least one value of ${\cal E}$ (for a fixed
$\gamma$) is available where $r^2_c$ would be exactly equal to $r^{gr}_c$.\\
\noindent
As $\gamma$ is increased, not much change is observed in the
general profile 
of ${\Delta}r^3_c$ except that $\left|{\Delta}r^3_c\right|_{max}$
{\it decreases} with {\it increase} in $\gamma$. The change in profile of 
${\Delta}r^4_c$  with increasing $\gamma$ is much more interesting. It is
observed that with increasing $\gamma$, not only does a zero appear in 
${\Delta}r^4_c$ vs ${\cal E}$ (for a fixed value of $\gamma$) curve
(which means that for some values of ${\cal E}$ and $\gamma$, 
$r^4_c$ can be exactly equal to $r^{gr}_c$), but also
the value of $\left|{\Delta}r^4_c\right|_{max}$ {\it decreases considerably} 
with an {\it increase}
in $\gamma$ and
as the flow approaches its purely non-relativistic limit,
we observe that the overall deviation produced by $\Phi_4$ in approximating
$r^{gr}_c$ is quite small in general 
(percentage deviation produced by 
$\Phi_4$ is found to be within the limit of 10$\%$ for 
purely non-relativistic
flow), which indicates that 
use of $\Phi_4$ is the 
{\it best possible} approximation of the Schwarzschild metric for
nonrelativistic flow (flow with high $\gamma$ value), regarding reproduction
of the sonic point for a flow with fixed ${\cal E}$ and $\gamma$. 
\begin{figure}
\vbox{
\vskip -4.8cm
\centerline{
\psfig{file=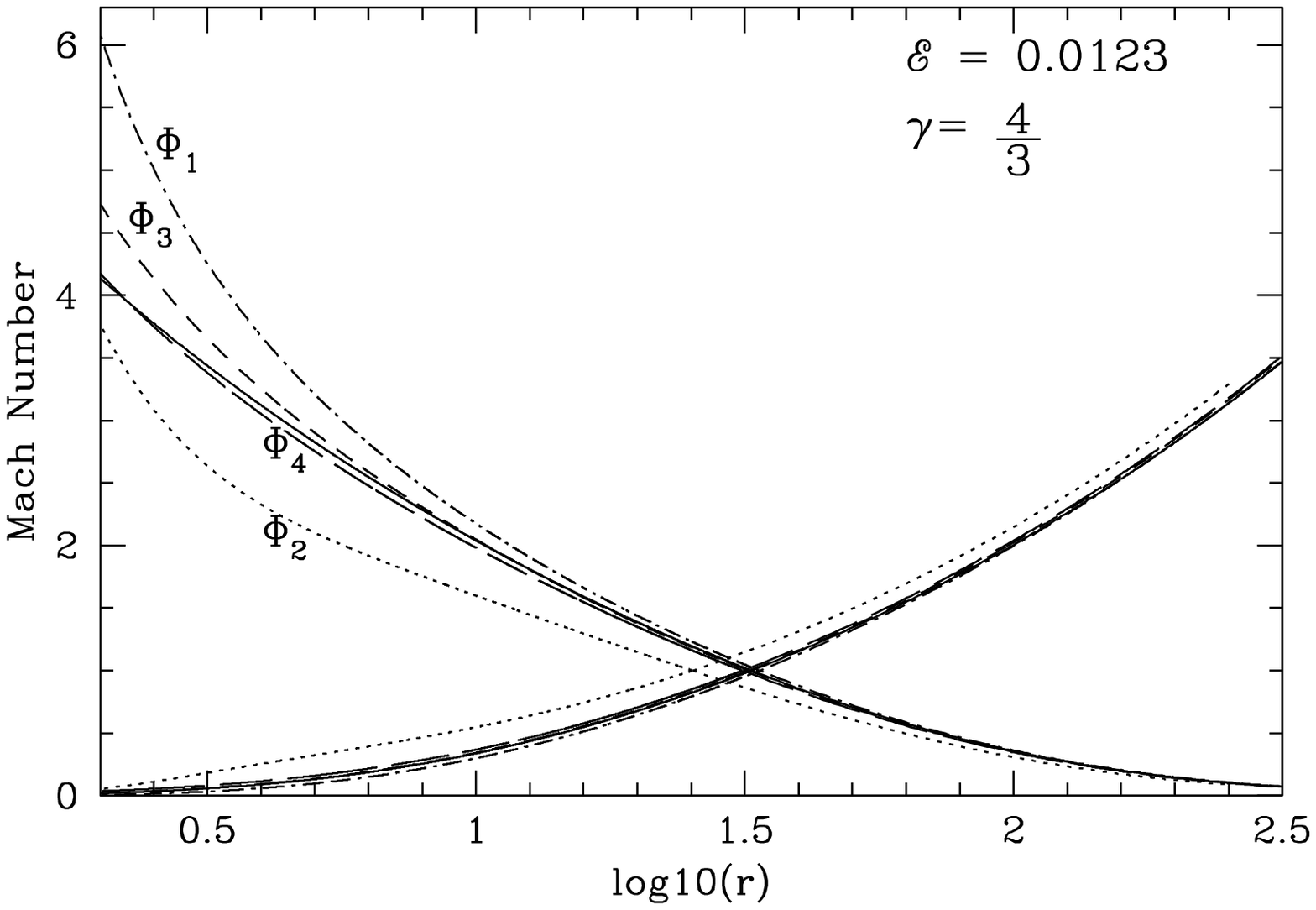,height=15cm,width=15cm}}}
\noindent {{\bf Fig. 3:}
Variation of Mach number $M$ (plotted along the Y axis)
with radial distance (plotted along X axis in logarithmic scale) for
accretion and wind for a fixed energy (${\cal E}=0.0123$) and 
adiabatic index ($\gamma=\frac{4}{3}$) of the flow. The solid lines represent
the variation obtained for FGRSF. For same ${\cal E}$ and $\gamma$, sonic
points $r^i_c$'s
obtained for flow in defferent pseudo-potentials and for FGRSF are all different
which is clearly observed from the figure. While the value of $r^{gr}_c$
is obtained as
$32.52 ~ r_g$, values of $r^1_c, r^2_c, r^3_c$ and $r^4_c$ are obtained as
$ 33.67~r_g,~25.27~r_g,~31.32~r_g$ and $32.12~r_g$
respectively where $r^i_c$'s are the sonic points for various $\Phi_i$'s.}
\end{figure}
\subsection{The integral curves of motion}
We have mentioned earlier that study of the integral curves,
i.e., the variation of Mach number of the flow $M$ with radial 
distance $r$
(measured from the accreting hole in units of $r_g$) is 
essential to investigate the `transonicity' of the flow. Let us first 
consider FGRSF. From eqn. 7(c), one can obtain the sonic point $r^{gr}_c$
for a fixed value of ${\cal E}$ and $\gamma$. The value
of $\left(\frac{du}{dr}\right)$
at $r^{gr}_c$ is then calculated using eqn. 7(d). One can numerically 
solve eqn. 5(a) and 5(b) simultaneously to obtain the value of the Mach 
number (and other dynamical quantities) as a function of radial distance
for a fixed value of ${\cal E}$ and $\gamma$. It is well known that for
spherically symmetric accretion onto a Schwarzschild black hole, two solutions
are obtained while solving the governing conservation equations, e.g., 
 equation for conservation of specific energy and from baryon number 
(or specific entropy)
conservation equation. One solution out of these two corresponds to the
accretion process and the other is for the wind. Using eqn. 6, 
it is easy to calculate the flow temperature at all points of the flow
(for accretion as well as for the wind branch). \\
\noindent
For accretion in pseudo potentials $\Phi_i$'s, the procedure is exactly the 
same. Sonic point $r^i_c$ is obtained by solving eqn. 10(c) for the $i$th
potential $\Phi_i$ for a fixed value of ${\cal E}$ and $\gamma$. 
 $\left(\frac{du}{dr}\right)_c$ for that particular
potential is then obtained by solving eqn. 10(d). Starting from $r^i_c$, 
eqn. 9(a) and eqn. 9(b) (alternatively, eqn. 9(a) and eqn. 9(c)) 
could be simultaneously solved (using eqn. 10(a)) to get the variation of
Mach number of the flow (as well as of other dynamical quantities) as a 
function of radial distance measured 
in units of $r_g$. The solution
for accretion as well as for wind branch can be obtained to investigate 
the `transonicity' of the flow by plotting Mach number $M$ as a function of 
radial distance $r$. Temperature of the flow $T^i$ (for any $i$th potential
$\Phi_i$) can easily be obtained by solving eqn. 6 for 
$a_i,p_i$ and $\rho_i$ for a fixed value of ${\cal E} $ and $\gamma$.\\
\noindent
In Fig. 3, we plot the variation of Mach number as a function of the 
radial distance 
(in logarithmic scale)
for FGRSF as well as for flows in all $\Phi_i$'s. The 
energy of the flow (for FGRSF and for flows in all $\Phi_i$'s) are kept
constant at a value $0.0123$ and value of $\gamma$ is taken to be $\frac{4}{3}$.
\begin{figure}
\vbox{
\vskip -5.8cm
\centerline{
\psfig{file=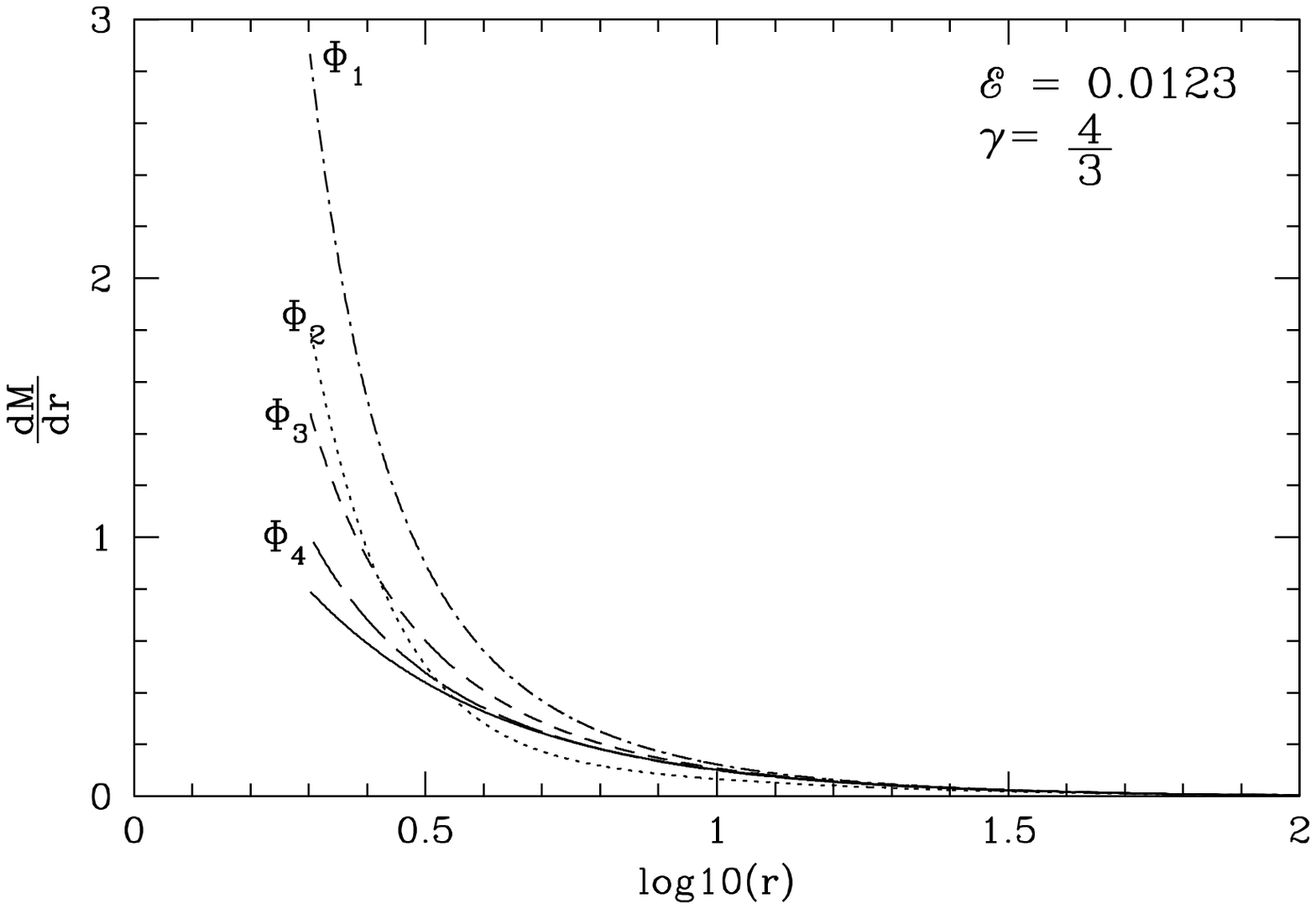,height=15cm,width=15cm}}}
\vbox{
\vskip -6.2cm
\centerline{
\psfig{file=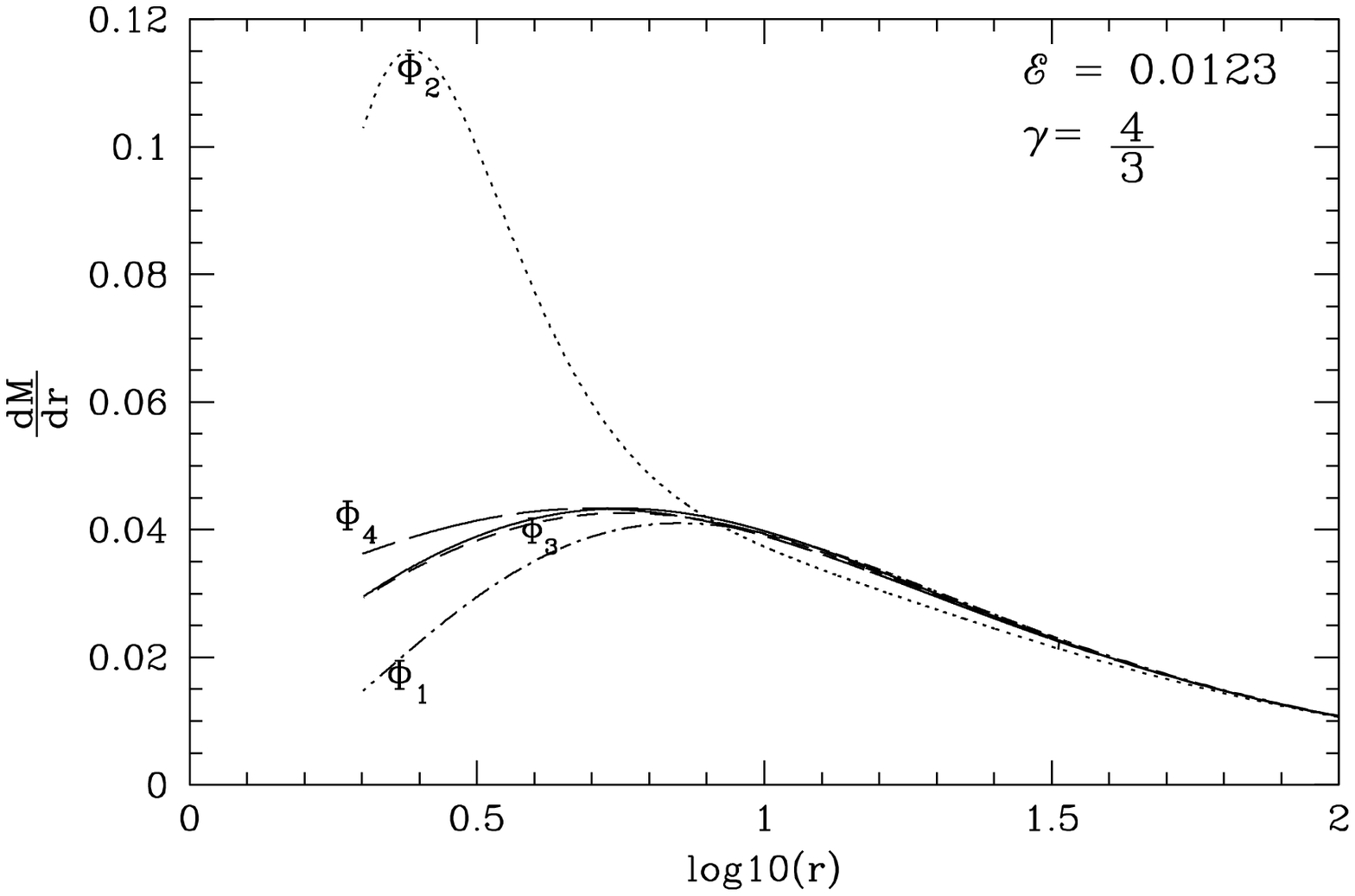,height=15cm,width=15cm}}}
\noindent {{\bf Fig.4 (a,b):}
Variation of the absolute value of the slope of
the Mach number $M$ (plotted on the Y axis as $\frac{dM}{dr}$) of
accretion (Fig. 4(a)) and wind (Fig. 4(b)) with radial distance
(plotted along X axis in logarithmic scale) measured from the accretor in units
of $r_g$. All input parameters are same as that of Fig. 3. Solid line
represents FGRSF. The slope in $M$ is found to increase towards the
accretor non-linearly and monotonically for the accretion branch and the space
rate of change of $M$ for $\Phi_1$ is clearly seen to be steepest while that
of $\Phi_4$ is observed to be the flattest. For wind, the change is not
monotonic; rather it produces a peak at the subsonic branch.}
\end{figure}
It is observed that sonic points of FGRSF $r^{gr}_c$, as well as flows in 
other potentials $r^i_c$ are {\it all different}, which is normally
expected (except for the very special value of ${\cal E}$ for which either the
${\Delta}r^2_c$ or ${\Delta}r^4_c$ vanishes, see Fig. 2(b) and 2(d)).
While the value of $r^{gr}_c$ is obtained as
$32.52 ~ r_g$, values of $r^1_c, r^2_c, r^3_c$ and $r^4_c$ 
are obtained as
$ 33.67~r_g,~25.27~r_g,~31.32~r_g$ and $32.12~r_g$
respectively. It is also to be noted that although the general profile
of the integral curves are almost same (except for 
flows in $\Phi_2$ where the 
change in curvature is slightly different to that of others, especially
for the supersonic accretion branch), the curves {\it do not} overlap in
general. 
The amount of deviation (from FGRSF) of integral curves drawn for
various $\Phi_i$'s is quite prominent for the supersonic branch
of accretion and the derivation is
observed to be not that significant 
(except for flows in $\Phi_2$) for corresponding
subsonic branches of accretion, as well as for the supersonic branch 
of winds. The deviation maximises with the decrease of radial distance 
from the accretion black hole. This is quite obvious because close to the
accretor, 
the general relativistic effects would be much more prominent thus the 
deviation would be higher. Also it is interesting to note that for the same 
energy ${\cal E}$ (and {\it even} for a fixed sonic point common to both
FGRSF and flows in all $\Phi_i$'s which may be obtained 
by properly tunning
the corresponding ${\cal E}$'s), the maximum value of the Mach number
$M_{max}$
of the flow in different $\Phi_i$'s
are very different.
One can observe from the figure that there exists a particular value
of $r$ (for a fixed ${\cal E}$ and $\gamma$) where the `crossing-over'
of ${\Phi_4}$ with FGRSF at the supersonic branch of the
accretion is quite distinct, that is, the
Mach number attained
for flows in $\Phi_4$ at a particular point is exactly 
equal to the Mach number produced by FGRSF. Also, the general profile for
the variation of Mach number of the flow in $\Phi_4$ is 
extremely close to the profile
obtained in FGRSF almost throughout the flow 
except very close ($r\sim8r_g$
in this case) to the black hole. However, no such mutual crossing over is observed
in between flows in various $\Phi_i$'s.\\
\noindent
If ${\Delta}T_{ke}$ and ${\Delta}T_{th}$ are the relative gain in the kinetic
energy and thermal energy of the flow respectively, it is easy to show 
that the change in Mach number ${\Delta}M$ could be approximated as:
$$
{\Delta}M~\propto~\sqrt{\frac{2}{\gamma-1}\left(\frac{{\Delta}T_{ke}}{{\Delta}T_{th
}}\right)}
\eqno{(13)}
$$
for spherical accretion in all pseudo potentials discussed here. Also it is 
easy to understand that the Mach number profile as well as the 
rate of change of Mach number at a particular point should be related to the 
form of the potential $\Phi_i$ used to study the flow (and in general to 
the metric used to describe the spacetime). Also $M^i_{max}$, the
maximum value of $M$ which could be attended for 
a particular set of ${\cal E}$ and $\gamma$ for flows in
any $i$th potential 
$\Phi_i$ (for the range of $r$ shown here),
should be related to the nature of the potential used to describe the
flow. It is observed that 
for $r~>~2r_g$, $M^i_{max}$ {\it anticorrelates} with 
${\cal S}_i$ (the `relative stiffness' of the $i$th potential) for
$i=1,3,4$ but {\it correlates} with ${\cal S}_i$ for $i=2$.\\
\noindent
As change in the Mach number is a result of a
mutual tug of war in between 
change in dynamical velocity and polytropic sound
speed (alternatively, in between the mechanical and thermal energy
of the flow, see eqn. 13), it might be interesting
to investigate the variation of the slope of the
Mach number (which might be
considered as the measure
of the degree of 
`transonicity' of the flow) with radial distance $r$ for
two different branches of solutions, namely, accretion and wind. In Fig. 
4(a) and 4(b) we plot the absolute value of $\left(\frac{dM}{dr}\right)$ 
as a function of $log10(r)$ for accretion (Fig. 4(a)) and wind 
(Fig. 4(b)) for FGRSF as well as for flows in various $\Phi_i$s. For 
accretion we see that $\left|\frac{dM}{dr}\right|$ increases nonlinearly and
{\it monotonically} as the flow comes closer to the black hole, 
which indicates that for all flows, instantaneous changes in kinetic energy of 
the flow is always {\it greater} than the instantaneous change in thermal 
energy, .i.e., ${\Delta}T_{ke}~>~{\Delta}T_{th}$ for all values of $r$, 
which is {\it not} the case for winds. For winds (see Fig. 4(b)), change of
$\left|\frac{dM}{dr}\right|$ is {\it  not} monotonic rather it 
{\it always shows a peak in the
subsonic branch} for FGRSF
as well as for flows in all $\Phi_i$'s. The appearance of the peak
for flows in pseudo potentials is due to the fact that from the close 
vicinity of the event horizon up to a certain distance $r_p$
($r^i_p~<~r^i_c$ and $r^{gr}_p ~<~r^{gr}_c$ always), the local gain in 
kinetic energy at any point is higher than the local thermal energy
gain. So up to $r_p, ~ {\Delta}T_{ke}~>~{\Delta}T_{th}$ but {\it after} 
$r_p$, ${\Delta}T_{ke}$ decreases and $\left|\frac{dM}{dr}\right|$ 
starts falling as the wind approaches to the sonic point. 
It is easy to 
understand that as ${\cal E}$ is kept constant throughout the flow (for accretion as well
as for the wind branch), the bulk motion acceleration of the flow is a 
continuous process throughout the accretion towards the black hole but
for winds the acceleration process dominates {\it only up to} $r_p$ and gets
the major part of the outward acceleration in the region bounded in 
$r_g~>r~>r^i_p$, though the exact physical reason behind this is not
clearly understood. However, it should be remembered that the above argument holds 
{\it only} for flows in pseudo potentials and {\it not} for FGRSF; because for
FGRSF, the total energy term can not be decoupled into various counterparts
with individual origin e.g., mechanical energy ${\cal E}_{ke}=\frac{u^2}{2}$
or thermal energy ${\cal E}_{th}=\frac{a^2}{\gamma-1}$ etc.\\
\noindent
It is clear from the figure
that the location of $r^1_p$ ($r_p$ for flows in $\Phi_1$)
is located furthest away from the black hole and $r^2_p$ is located 
closest to the black hole with the following sequence:
$$
r^1_p~<~r^3_p~<~r^{gr}_p~<~r^4_p~<r^2_p
\eqno{(14)}
$$
It is also observed that with increase of $\gamma$, the sequence in eqn. (14) 
is maintained but the location of all $r_p$s moves {\it towards} the
black hole. Thus, the regions from where the wind (in various $\Phi_i$'s) 
is accelerated, moves closer to the hole as the flow approaches to its 
non-relativistic limit. 
If we define $\Delta{\left(\frac{dM}{dr}\right)}^i$ to be:
$$
\Delta{\left(\frac{dM}{dr}\right)}^i=\frac{
{{\left(\frac{dM}{dr}\right)}^{i}}-{\left(\frac{dM}{dr}\right)}^{gr}
}
{{\left(\frac{dM}{dr}\right)}^{gr}}
\eqno{(15)}
$$
then one can show
that for a fixed value of ${\cal E}$ and $\gamma$, the 
following sequence are maintained for various $\Phi_i$s as:
$$
{\left|\Delta{\left(\frac{dM}{dr}\right)}^3\right|}_{max}~<~
{\left|\Delta{\left(\frac{dM}{dr}\right)}^4\right|}_{max}~<~
{\left|\Delta{\left(\frac{dM}{dr}\right)}^2\right|}_{max}~<~
{\left|\Delta{\left(\frac{dM}{dr}\right)}^1\right|}_{max}
$$
for accretion where $
{\left|\Delta{\left(\frac{dM}{dr}\right)}^i\right|}_{max}$
stands for the maximum value of $\Delta{\left(\frac{dM}{dr}\right)}^i$. 
Whereas for wind it is observed that:
$$
{\left|\Delta{\left(\frac{dM}{dr}\right)}^4\right|}_{max}~<~
{\left|\Delta{\left(\frac{dM}{dr}\right)}^3\right|}_{max}~<~
{\left|\Delta{\left(\frac{dM}{dr}\right)}^1\right|}_{max}~<~
{\left|\Delta{\left(\frac{dM}{dr}\right)}^2\right|}_{max}
$$
Also note that for the accretion branch all 
$\Delta{\left(\frac{dM}{dr}\right)}^i$ change sign except 
$\Delta{\left(\frac{dM}{dr}\right)}^1$.\\
\begin{figure}
\vbox{
\vskip -5.0cm
\centerline{
\psfig{file=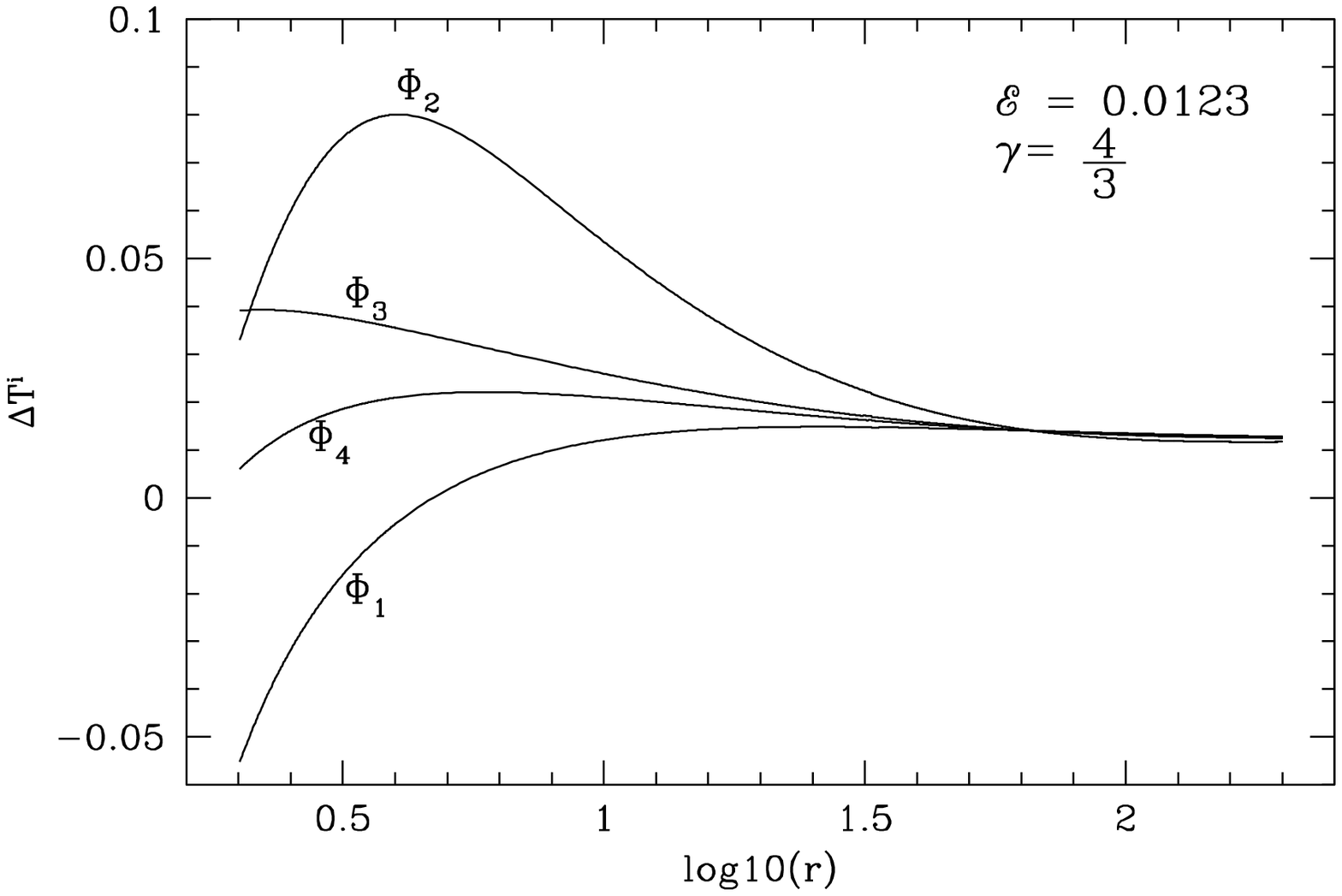,height=15cm,width=15cm}}}
\noindent {{\bf Fig. 5:}
Variation of the temperature deviation (${\Delta}T^i$s) of
the accretion flow with radial distance for ultra relativistic
($\gamma=\frac{4}{3}$) flow in various $\Phi_i$s. While the difference
in flow temperature in between any $T^i$ 
(flow temperature for a particular $i$th potential
$\Phi_i$) and $T^{gr}$
(flow temperature for FGRSF), is plotted along
the Y axis, the distance from the 
central
accretor (in logarithmic scale in units of $r_g$) is plotted along X axis.
The maximum amount of deviation is obtained close to the accretor,
which is
expected. It is clearly observed from the figure that use of $\Phi_4$ and
(except very close to the black hole) $\Phi_1$ are the
best approximation of the 
FGRSF flow temperature. Almost the same sort of deviation
is found for the wind branch (not shown in the figure).}
\end{figure}
\subsection{Temperature profile of the flow}
\noindent
One can easily show that at any point of the flow, 
the flow temp $T^i$ for flow in any particular 
$\Phi_i$s can be re-written as:
$$
T^i=\frac{{\mu}m_p}{{\gamma}{\kappa}}\left\{\frac
{\left(2\gamma-1\right)\left({\cal E}-{\Phi_i}\right)}
{2+M^i\left(\gamma-1\right)}
\right\}
\eqno{(16)}
$$
where $M^i$ is the corresponding Mach number for the $i$th pseudo potential.
 As it has been stated that $M^i$ can be calculated at each point of the flow, 
one can compute $T^i$ from knowledge of $M^i$s using eqn. (16). After defining ${\Delta}T^i$ to be:
$$
{\Delta}T^i=\frac{T^i-T^{gr}}{T^{gr}}
\eqno{(17)}
$$
 where $T^{gr}$ could be computed using eqn. 6, one can plot ${\Delta}T^i$
as a function of radial distance (in units of $r_g$) to investigate 
which $\Phi_i$ produces the
corresponding $T^i$ closest to $T^{gr}$ at a 
particular radial distance and for a fixed value of the
${\cal E}$ and $\gamma$.
In Fig. 5. we plot ${\Delta}T^i$ vs $log10(r)$ to demonstrate the dependence 
of ${\Delta}T^i$ on radial distance. It is clear from the figure that except
for 
$T^1$, all other $T^i$s are always {\it higher} than $T^{gr}$,
which indicates
that all $\Phi_i$s except $\Phi_1$ produce {\it hotter flow} than
FGRSF in general. 
 For $\Phi_1$, very close to the hole the flow is {\it cooler} compared
to FGRSF up to a certain distance,
after which $\Phi_1$ also produces 
$T^1$ {\it higher} to $T^{gr}$. The point at which $T^1$ becomes equal to 
$T^{gr}$, gets
shifted away from the black hole with increasing $\gamma$,
keeping ${\cal E}$ fixed. Also,
it is observed that $\left|{\Delta}T^i\right|_{max}$
{\it increases} with increasing $\gamma$. From Fig. 5 it is clear that 
throughout its way from infinity to the event horizon, $\Phi_1$ and $\Phi_4$
produces very good approximation of FGRSF regarding calculation of
flow temperature. However, after a certain point very close to the 
black hole (after the crossing the point where $T^1=T^{gr}$), 
 $\Phi_1$ produces a sudden and sharp deviation of the
 flow temperature from 
FGRSF, physical reason for which is not clearly understood. 
More or less the 
same sort of profile in the 
${\Delta}T^i$ vs $log10(r)$ curve also
is observed for wind branch.
\section{Conclusion}
In this paper we 
have solved a set of algebraic and differential equations
governing various dynamical and thermodynamic behaviouars
of Bondi (1954) type accretion in a full Schwarzschild metric
as well as for motion under a number of `pseudo-Schwarzschild' 
potentials, to examine the suitability in
application of those 
potentials in investigating spherically symmetric
transonic accretion onto a non- rotating black
hole. We have shown that though the potentials
discussed here were originally proposed to mimic the
relativistic effects manifested in the disc accretion,
it is quite reasonable to use most of the
potentials in studying various dynamical as well as thermodynamic
quantities
for spherical accretion. Also, we have shown that
depending on the chosen regions of parameter space
spanned by specific energy ${\cal E}$ and adiabatic index
$\gamma$ of the flow, one potential may be
important than others and we could identify which
potential is the best approximation for FGRSF for what
values of ${\cal E}$ and $\gamma$.
We have restricted ourselves to the
study of simple polytropic flows only.
However, the
validity of using all these $\Phi_i$s discussed here can easily
be examined for isothermal accretion and wind as well as for flows
with other equations of state. Work in this direction is 
reported elsewhere (Sarkar \& Das, 2001).\\
\noindent
It is observed that among all pseudo
potentials, $\Phi_1$ (potential proposed by Paczy\'nski and Wiita
(1980)) and $\Phi_4$ (one of the potentials proposed by
Artemova et. al. (1996)) are in general the best in the
sense that they provide very reasonable approximation
to the full general relativistic solution. While $\Phi_1$
is the best approximation for ultra-relativistic flow,
$\Phi_4$ happens to be the best approximation as the flow
tends to be fully non relativistic, i.e,
$\gamma$ tends to have the
value $\frac{5}{3}$. Also we see that there are certain cases for
which one or more of the pseudo potentials may give the
exact match with FGRSF for a particular value of ${\cal E}$ or
$\gamma$
(for a fixed $r$) in finding some dynamical ($r_c, M$ etc.)  or
thermodynamic (flow temperature $T$, for example) quantity.\\
\noindent
It is worth mentioning that as long as one is not
interested in astrophysical processes extremely close
(within $1-2~r_g$) to a black hole horizon, one may safely
use the `pseudo' potentials discussed here to study
spherically symmetric accretion on to a Schwarzschild
black hole with the advantage that use of these
potentials would simplify calculations by allowing one
to use some basic features of flat geometry
(additivity of energy or de-coupling of various
energy components, i.e., thermal ($\frac{a^2}{\gamma-1}$)
Kinetic ($\frac{u^2}{2}$) or
gravitational ($\Phi$) etc.) which is not possible for
calculations in a purely Schawarzschild metric. Also, one
can study more complex many body problems such as
accretion from an ensemble of companions or overall
effeciency  of accretion onto an ensemble of black holes
in a galaxy  or  for studying numerical hydrodynamic or
magnetohydrodynamic flows
around a black hole etc. as simply as can be done in a 
Newtonian framework, but with far better
accuracy. However, one should be careful in using these
potentials to study spherically symmetric accretion
because of the fact that none of the potentials discussed here
are `exact' in a sense that they are not directly
derivable from the Einstein equations. 
These potentials
could only be used to obtain more
accurate correction terms over and above the pure
Newtonian results and any `radically' new results
obtained using these potentials should be cross-checked
very carefully with the exact general relativistic theory.\\
\noindent
Although the theory of disc accretion has priority
over spherical accretion because of the fact that
accretion discs describe more realistic situations
found in nature, it is not unreasonable to
concentrate on spherical accretion because for
certain cases, that may be quite useful and 
use of these potentials makes a complicated problem 
simpler to
study. For example, for a supermassive black hole 
immersed in intergalactic space in such a way that
matter falling on to it has negligible intrinsic
angular momentum, the accretion (at least
close to the hole) is quasi spherical and transonic spherical 
accretion might be a good approximation to mimic the
situation. Same sort of approximation is valid when an
accreting black hole is embedded in a number of donor
stars (or star clusters)  where the angular momentum of the stars are
randomly oriented in such a way that the vector sum
of the intrinsic angular momentum carried by the
accreting matter as a whole may be quite negligible, so as to 
make Bondi-type accretion a good approximation. In fact, a
number of recent works (Coker \&
Markoff 2001 and references therein,
Das, 1999, 2000, 2001a,b,
Toropin, et. al, 1999, Kovalenko \& Eremin 1998,
Titarchuk, et. al. 1996, 1997,
Wang \& Sutherland 1997, Zampieri, et. al. 1996,
Yim \& Park 1995, Markovic 1995, 
Tsuribe, et. al. 1995, Kazhdan \& Murzina 1994,
Fortner 1993) still deal with spherical
accretion to investigate some basic astrophysical
processes the black holes and neutron stars. So we believe that
work presented in this paper is relevant and will be useful
in investigation of
various aspects of accretion and wind around 
non-rotating and slowly rotating compact objects.
\begin{acknowledgements}
We are grateful to Prof. P. J. Wiita and Prof. I. Novikov for useful
discussions.
\end{acknowledgements}

\end{document}